# Scattering-asymmetry control with ultrafast electron wave packet shaping

Yuya Morimoto[1,2,3]*, Peter Hommelhoff[2,4] and Lars Bojer Madsen[5†]

*[1] Pioneering Research Institute (PRI) and RIKEN Center for Advanced Photonics (RAP), RIKEN, 2-1 Hirosawa, Wako, Saitama, 351-0198, Japan*

*[2] Laser Physics, Department of Physics, Friedrich-Alexander-Universität Erlangen-Nürnberg (FAU), Staudtstraße 1, 91058, Erlangen, Germany*

*[3] Department of Nuclear Engineering and Management, School of Engineering, The University of Tokyo, 7-3-1 Hongo, Bunkyo-ku, Tokyo, 113-8656, Japan*

*[4] Department of Physics, Ludwig-Maximilians-Universität München, 80799 München, Germany*

*[5] Department of Physics and Astronomy, Aarhus University, Ny Munkegade 120, 8000 Aarhus C, Denmark*

*yuya.morimoto@riken.jp

†bojer@phys.au.dk

**Scattering of a tightly focused electron beam by an atom forms one of the bases of modern electron microscopy. A fundamental symmetry breaking occurs when the target atom is displaced from the beam center. This displacement results in a deflection of the beam and an asymmetric angular distribution of the scattered electrons. Here we propose a concept to control the sign and magnitude of the scattering asymmetry by shaping the incident high-energy electron wave packet in momentum space on the atto- to picosecond time scale. The shaping controls the ultrafast real-space dynamics of the wave packet, shifting the balance between two competing contributions of the impact-parameter-dependent quantum interference and the momentum distribution of the wave packet on the target.**

**We find a strong sensitivity of the elastic scattering on the wave packet properties, an effect that will allow wave-packet and target characterization in ultrafast electron microscopy.**

## 1. Introduction

The conventional theoretical framework describing electron-atom scattering involves a plane-wave description of the incoming projectile electron at large distances as part of the asymptotic boundary condition [1–3]. This means that the incoming beam is assumed to be infinitely large transversally and of infinite duration. In other words, the incident electron is assumed to be monochromatic with a single fixed longitudinal momentum. Electron scattering cross sections calculated with these assumptions form the foundation of various research fields [1–3].

In contrast, two technical advances allow producing electron beams that cannot be described asymptotically as plane waves. One is the ability to generate an electron beam with a very tight focus. State-of-the-art scanning transmission electron microscopes (STEMs) provide electron beams of sub-Angstrom size [4–8]. In this case, because of the position-momentum uncertainly, a transversal momentum distribution needs to be considered [9–12]. When the target atom is displaced from the center of the focused beam, the beam is deflected due to electromagnetic field of the atom [4–8] and an asymmetry of total signals appears on a detector. Quantum mechanically, the deflection can be interpreted as the interference between the transmitted and elastically-scattered waves [9,11,12]. In addition, an asymmetry appears in the angular distribution of the scattered electrons themselves [13,14]. These two asymmetries are the focus of this work. The other notable development reported recently is the coherent manipulation of electron beams by light waves [15–25]. Electron pulses as short as hundreds of attoseconds were experimentally obtained [17,18,20–26]. The use of optically modulated electron pulses for controlled excitation of a two-level system [27–30] or coherent photon generation [31–34] were proposed. The confinement of electrons within an extremely short duration is accompanied by a broad longitudinal momentum distribution. If these two techniques, spatial focusing and temporal shaping, are combined, which should be within reach considering the rapid advances of ultrafast electron microscopes [35], attosecond and sub-Angstrom scale electron microscopy will be possible. A comprehensive understanding of scattering of electron wave packets shaped in three-dimensional space is, however, currently lacking.

In this work, we remedy this situation by theoretically investigating fundamental physical quantities in elastic scattering of electron wave packets shaped in time and space. We consider two types of Gaussian wave packets having atto- to picosecond durations. One has a Gaussian distribution in kinetic energy and the other in the longitudinal momentum. We show that the two types of wave packets exhibit completely different features when the symmetry of the system is broken due to displacement of the target atom from the beam center. We also show that the angular asymmetry of elastic scattering itself, which has not been used for atomic imaging, contains rich information about the properties of ultrashort electron wave packets. We explain the observations based on quantum-mechanical interference originating from the displaced target and the spatial-momentum overlap between the target and the wave packet occurring on sub-fs time scales. Our findings facilitate the control of the symmetry, strength, phase, and timing of electron-matter interaction by optical electron-beam manipulation in three dimensions.

## 2. Theory

The physical system under consideration is illustrated in Fig. 1(a). A non-relativistic ultrashort electron wave packet propagating along the *z*-axis is elastically scattered by a target located at $z = 0$, where the wave packet is transversally focused. We assume a spatially fixed target, for example, an atom in an ultracold quantum gas [36] or a defect atom in a two-dimensional crystal [37,38]. The *x*-axis describes the displacement of the target from the beam center. Accordingly, the impact parameter is expressed as $\boldsymbol{b} = b_x\hat{\boldsymbol{x}}$, where $\hat{\boldsymbol{x}}$ is the unit vector along the *x* axis. The momenta of the electrons before and after the interaction with the target, $\hbar\boldsymbol{k}_i$ and $\hbar\boldsymbol{k}_f$, are described by their absolute values $\hbar k_i$ and $\hbar k_f$, and polar and azimuthal angles ($\theta_i$, $\varphi_i$) and ($\theta_f$, $\varphi_f$), see Figs. 1(a) and (c). The time-dependent electron wave packet in real space is given by

$$\psi_e(\boldsymbol{r}, t) = \frac{1}{(2\pi)^{\frac{3}{2}}} \int d\boldsymbol{k}_i\, a_e(\boldsymbol{k}_i) \exp\left(\mathrm{i}\boldsymbol{k}_i \cdot \boldsymbol{r} - \frac{\mathrm{i}E_i t}{\hbar}\right), \tag{1}$$

where $E_i = \hbar^2 k_i^2/(2m_e)$ is the kinetic energy and $a_e(\boldsymbol{k}_i)$ is the momentum space wavefunction. Outside the range of the target potential, this wavefunction satisfies the time-dependent Schrödinger equation with any choice of $a_e(\boldsymbol{k}_i)$. The probability to find an electron at the final momentum $\boldsymbol{k}_f$ after the scattering is derived from the time-dependent *S*-matrix theory [39–41]. By neglecting the exchange interaction, which is valid for high-energy electrons (10 keV in this work), and assuming that the target is much heavier than the projectile electrons, the probability is obtained as [40,41]

$$P_{\boldsymbol{b}}(\boldsymbol{k}_f) = I_{unsc}(\boldsymbol{k}_f) + I_{int,\boldsymbol{b}}(\boldsymbol{k}_f) + I_{sc,\boldsymbol{b}}(\boldsymbol{k}_f), \tag{2}$$

where

$$I_{unsc}(\boldsymbol{k}_f) = \left|a_e(\boldsymbol{k}_f)\right|^{\mathbf{2}}, \tag{3}$$

$$I_{int,\boldsymbol{b}}(\boldsymbol{k}_f) = -\frac{1}{\pi}\mathrm{Im}\left[a_e^*(\boldsymbol{k}_f)k_f \int d\widehat{\boldsymbol{k}}_i\, a_e(k_f\widehat{\boldsymbol{k}}_i)e^{-\mathrm{i}(\boldsymbol{k}_f - k_f\widehat{\boldsymbol{k}}_i)\cdot\boldsymbol{b}} f_{elas}\,(k_f\widehat{\boldsymbol{k}}_i, \boldsymbol{k}_f)\right], \tag{4}$$

and

$$I_{sc,\boldsymbol{b}}(\boldsymbol{k}_f) = \frac{1}{4\pi^2}\left|k_f \int d\widehat{\boldsymbol{k}}_i\, a_e(k_f\widehat{\boldsymbol{k}}_i)e^{\mathrm{i}k_f\widehat{\boldsymbol{k}}_i\cdot\boldsymbol{b}} f_{elas}\,(k_f\widehat{\boldsymbol{k}}_i, \boldsymbol{k}_f)\right|^2, \tag{5}$$

respectively, where $\widehat{\boldsymbol{k}}_i = \boldsymbol{k}_i/k_i$ and $k_f = k_i$. The elastic scattering amplitude $f_{elas}\,(k_f\widehat{\boldsymbol{k}}_i, \boldsymbol{k}_f)$ is obtained by the nonperturbative calculations [42] and the phase shift due to the Columbic potential of the target is considered. The derivation of these equations is given in [40]. Equation (2) shows that the total probability is given by the sum of three terms. $I_{unsc}(\boldsymbol{k}_f)$ represents the intensity profile of the un-scattered part, $I_{int,\boldsymbol{b}}(\boldsymbol{k}_f)$ the interference between the un-scattered and scattered waves, and $I_{sc,\boldsymbol{b}}(\boldsymbol{k}_f)$ the profile of the scattered part. We note that these three contributions cannot be distinguished experimentally, but their relative contributions change with scattered angles. In general, $I_{unsc}(\boldsymbol{k}_f)$ has the narrowest angular profile followed by $I_{int,\boldsymbol{b}}(\boldsymbol{k}_f)$ and $I_{sc,\boldsymbol{b}}(\boldsymbol{k}_f)$ [40]. In this work, we consider a hydrogen atomic target which can be displaced along the $x$ axis in Fig. 1(a), and electrons are detected by a two-dimensional detector without energy resolution as in STEM. Therefore, a signal profile on the detector is given by

$$\tilde{P}_{b_x}(\theta_f, \varphi_f) = \int dk_f\, k_f^2 P_{b_x}(\boldsymbol{k}_f). \tag{6}$$

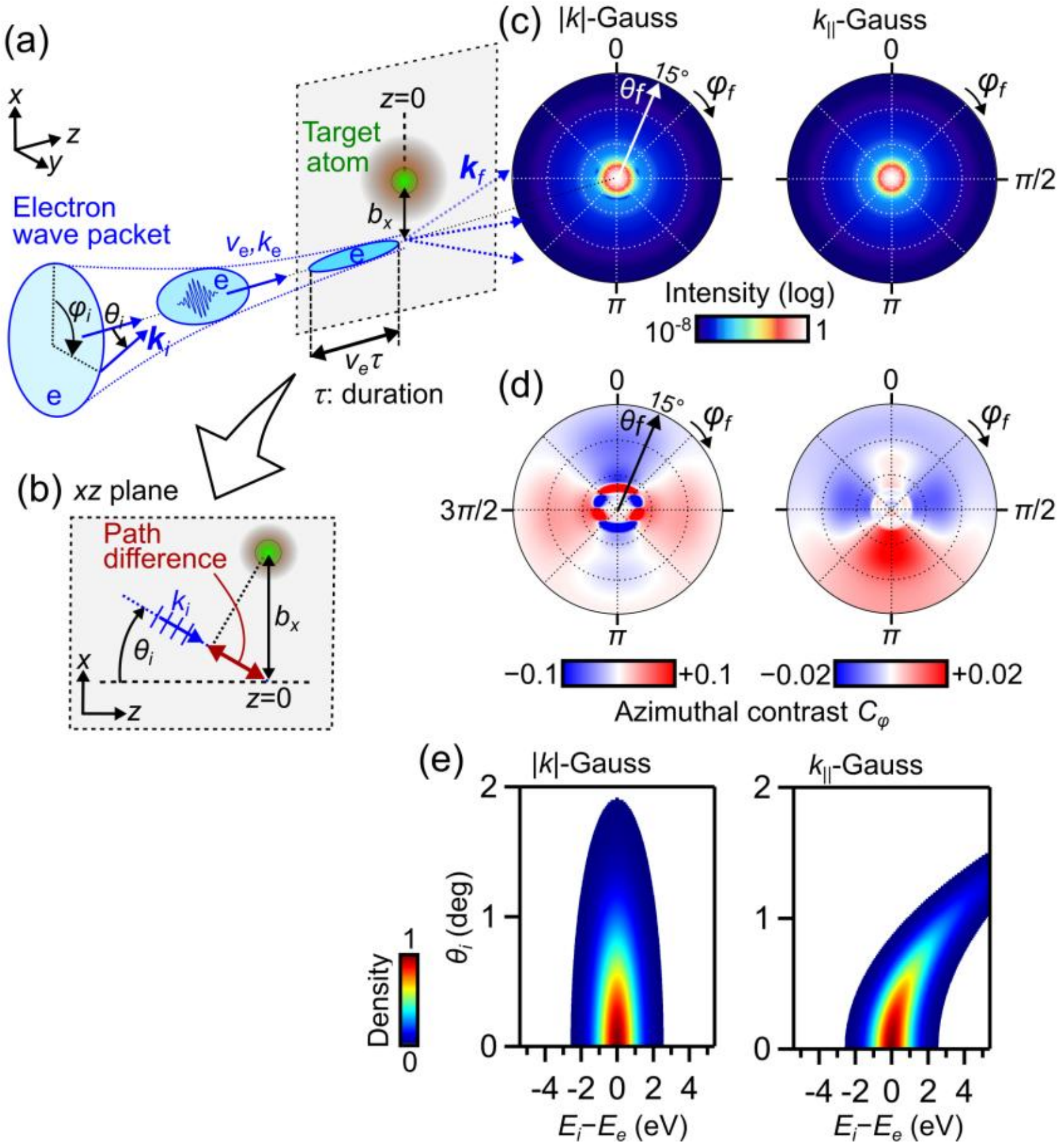

Fig 1. Scattering of a 3D-shaped ultrashort electron wave packet. (a) The spatially focused electron pulse is scattered by a target displaced from the beam center. The target represents an atom in an optically trapped atomic array or in a two-dimensional crystal. (b) Illustration of the path length difference, leading to the $\boldsymbol{b}$-dependent phase of Eq. (5). (c) Polar plots of the total signals [Eq. (6)] calculated at $\tau = 1$ fs and $b_x = 2$ Å. The radius corresponds to the final angle $\theta_f$. (d) Polar plots of the azimuthal contrast [Eq. (9)] calculated at the same condition for panel (c). (e) Energy—initial-angle ($\theta_i$) distributions of the two types of wave packets. The highest density in each plot is normalized to one.

In the simulations below, we assume a circular detector with a maximal detection angle of $\theta_f$ = 15 deg (262 mrad) which is close to the typical maximal angle (~200 mrad) for the high-angle annular dark-field (HAADF) STEM imaging [43]. Also, this angle is equivalent to 54 mrad (3 deg) at the kinetic energy of 200 keV in terms of the momentum transfer.

We consider two types of axially symmetric 10-keV electron-beam wave packets. The axial symmetry allows us to describe the momentum-space wavefunction $a_e(\boldsymbol{k}_i)$ by $(k_i, \theta_i)$. Their momentum-space densities for 1-fs pulses are plotted in Fig. 1(e). The first one (right panel) is defined as

$$a_e(k_i, \theta_i) = \frac{1}{(8\pi^3)^{\frac{1}{4}}\sqrt{\sigma_\parallel}\sigma_\theta k_e}\exp\left(-\frac{(k_\parallel - k_e)^2}{4\sigma_\parallel^2}\right)\exp\left(-\frac{\sin^2\theta_i}{4\sigma_\theta^2}\right)\exp\left[\mathrm{i}\phi_{prop}(t_p)\right], \tag{7}$$

where $k_e$ is the momentum magnitude at the energy of $E_e$ = 10 keV, $k_\parallel = k_i\cos\theta_i$ is the longitudinal momentum, $\hbar\sigma_\parallel$ the root-mean-square (rms) longitudinal momentum width and $\phi_{prop}(t_p)$ is the phase shift accounting for the pulse dispersion associated with the free-space propagation duration $t_p$. The parameter $t_p$ controls the $z$-position of the longitudinal (i.e., temporal) focusing [41]. By defining the phase as $\phi_{prop}(t_p) = t_p(k_\parallel v_e - \hbar k_\parallel^2/(2m_e))$ where $v_e$ is the group velocity of the wave packet, the transversal focus occurs always at $z = 0$ where the target is located at $t = 0$ [41]. With nonvanishing phase $\phi_{prop}(t_p)$, there is a difference in the transversal and longitudinal (i.e., temporal) focus points. In this work, we set the parameter as $t_p = 0$, that is, the longitudinal size (i.e., the pulse duration) is the shortest at $z = 0$. In real space, the rms and FWHM longitudinal sizes are $1/(2\sigma_\parallel)$ and $\sqrt{2\ln(2)}/\sigma_\parallel$, respectively, and the corresponding FWHM duration is given by $\tau = \sqrt{2\ln(2)}/(\sigma_\parallel v_e)$. $\sigma_\theta$ is the rms angular width and throughout we use $\sigma_\theta$ =10 mrad, a typical value for STEM [43]. The corresponding rms spot size is $1/(2k_e\sigma_\theta)$ = 1 Å. Since this wave packet has a Gaussian distribution in the longitudinal momentum, we call it a $k_\parallel$-Gauss wave packet. The wavefunction is normalized, $\int |a_e(k_i, \theta_i)|^2 \, d\boldsymbol{k}_i = 1$.

The second wave packet [left panel of Fig. 1(e)] is defined as

$$a_e(k_i, \theta_i) = \frac{1}{(8\pi^3)^{\frac{1}{4}}\sqrt{\sigma_\parallel}\sigma_\theta k_e} \exp\left(-\frac{(k_i - k_e)^2}{4\sigma_\parallel^2}\right) \exp\left(-\frac{\sin^2\theta_i}{4\sigma_\theta^2}\right) \exp[\mathrm{i}\phi_{prop}(t_p)]. \tag{8}$$

Since this wave packet has a Gaussian distribution in the momentum magnitude $k_i$, we call it a $|k|$-Gauss wave packet. As for the above wave packet type, we assume $t_p = 0$. In the limit $\sigma_\parallel \to +0$ (i.e., $\tau \to +\infty$), it resembles the monochromatic but transversally focused beams in electron microscopes. Examples of the signals on the detector $\tilde{P}_{b_x}(\theta_f, \varphi_f)$ [Eq. (6)] with the two types of wave packets at $\tau$ =1 fs and $b_x$= 2 Å are shown in Fig. 1(c), where the radius corresponds to the final angle $\theta_f$.

We consider wave-packet durations $\tau$ from 1 as to 1 ps. In the limit $\sigma_\parallel \to +\infty$ (i.e., $\tau \to +0$), the two wave packets are nearly identical. On the other hand, in the limit $\sigma_\parallel \to +0$ (i.e., $\tau \to +\infty$), they give completely different scattering probabilities and asymmetries, as we show below. While the momentum-space distribution of the $|k|$-Gauss wave packet [Eq. (8)] is a two-dimensional Gaussian [left panel of Fig. 1(e)], that of the $k_\parallel$-Gauss wave packet [Eq. (7)] forms an arc due to the energy-angle correlation [right panel of Fig. 1(e)]. The arc shape becomes clear when the mean value of total momentum at $\theta_i = \sigma_\theta$ deviates from that at $\theta_i = 0^\circ$ by more than the variation $\sigma_\parallel$, which is $\sqrt{k_e^2 + k_e^2\sin^2\sigma_\theta} - k_e \geq \sigma_\parallel$. This leads to $\tau \geq 2\sqrt{2\ln(2)}/(k_e v_e \sin^2\sigma_\theta)$ =0.8 fs. Therefore, a significant difference in the two wave-packet types is expected to appear at the duration of sub-femtosecond or longer ($\tau \geq$ fs). The two wave-packet types can in principle be produced with existing experimental techniques, see section 5. Experimental feasibility below.

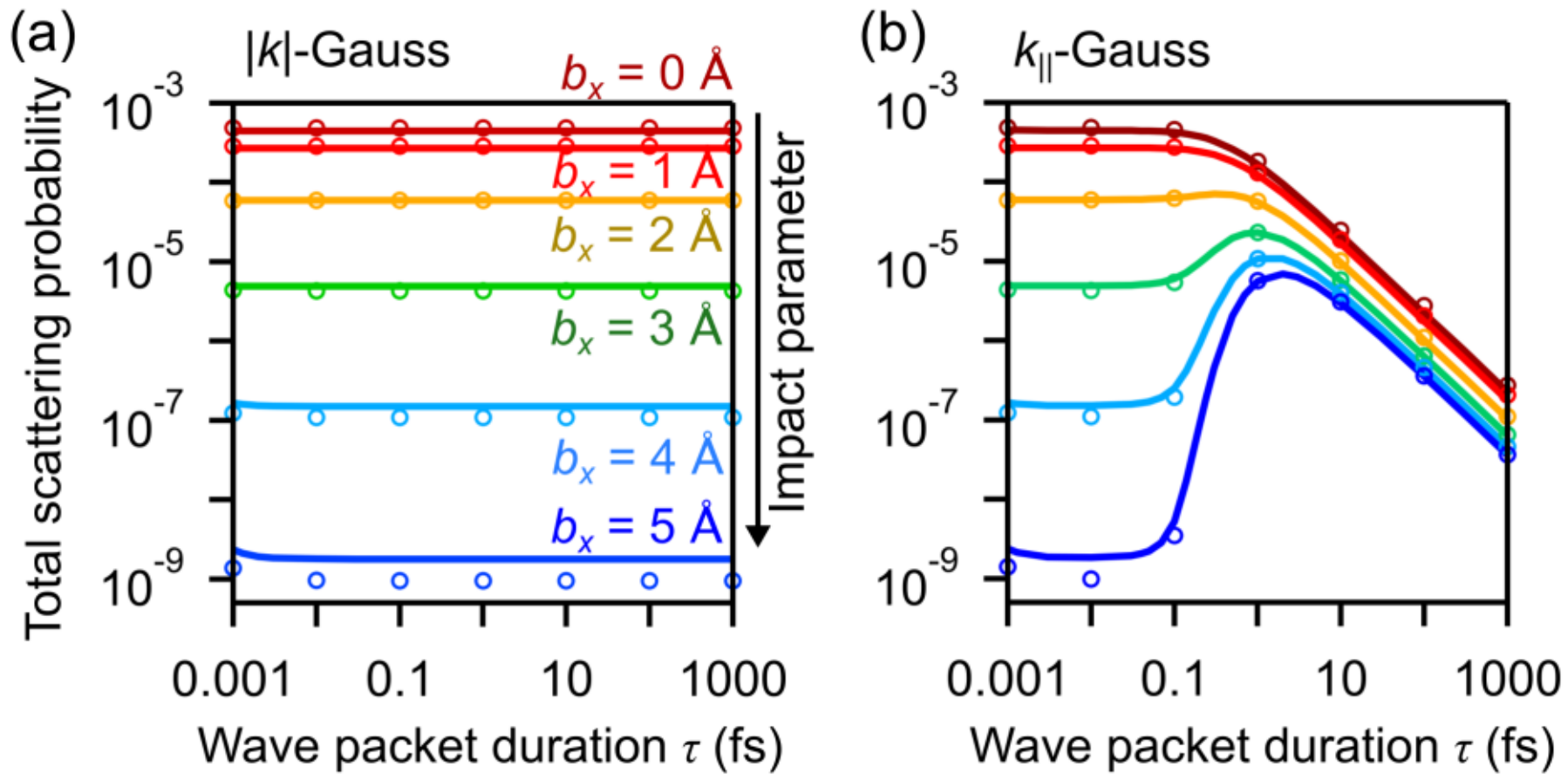

Fig 2. Total scattering probability by a hydrogen atom. (a) Total elastic scattering probabilities as a function of wave packet duration and impact parameter (solid curves) for the $|k|$-Gauss wave packet. (b) Total elastic scattering probabilities for the $k_{\|}$-Gauss wave packet. Circles represent scattering probabilities estimated from the normalized flux at the position of the target [Eq. (12)].

## 3. Results

We first compare the total elastic scattering probabilities $I_{sc,total}(b_x) = \int I_{sc,\boldsymbol{b}}(\boldsymbol{k}_f) d\boldsymbol{k}_f$ for the two wave packet types in Fig. 2 as a function of their duration $\tau$. Impact parameters $b_x$ from 0 Å to 5 Å are considered. We note that the integral of the total signals on the detector, $P_{\boldsymbol{b}}(\boldsymbol{k}_f)$ in Eq. (2), over $\boldsymbol{k}_f$ is unity because the integral of $I_{unsc}(\boldsymbol{k}_f)$ [Eq. (3)] is unity due to the normalization and that of $I_{int,\boldsymbol{b}}(\boldsymbol{k}_f)$ [Eq. (4)] is equal to $-I_{sc,total}(b_x)$ due to the flux conservation, i.e., optical theorem [40]. The $|k|$-Gauss wave packet (left) results in $I_{sc,total}(b_x)$ being nearly independent of the wave-packet duration. However, the $b_x$-dependence is significant: $I_{sc,total}(b_x)$ at $b_x = 5$ Å is five orders of magnitude lower than that at $b_x = 0$ Å with a beam of ~1 Å rms spot size. This high contrast provides the ability to probe atoms with high spatial resolution, which is the physical background of the atomic-resolution STEM [43,44]. In contrast, the total elastic-scattering probability of the $k_{\|}$-Gauss wave packet (right panel) strongly depends on the wave-packet duration $\tau$. For this pulsed beam, we identified three regimes. At very short durations of $\tau < 0.1$ fs, the results are nearly identical to those of the $|k|$-Gauss. This is because the two types of wave packets cannot be distinguished in the limit $\sigma_{\|} \rightarrow +\infty$ (i.e., $\tau \rightarrow +0$). At the transient regime from 0.1 fs to 10 fs, the scattering probability drastically changes, especially for the larger impact parameters considered. In the long pulse regime ($\tau$ >10 fs), the scattering probability decreases monotonically. The origin of the duration dependence is the local densities of the wave packets at the position of the target [open circles in Fig. 2(a),(b)], see below and Fig. 4.

To examine how the two-dimensional signals on the detector $\tilde{P}_{b_x}(\theta_f, \varphi_f)$ change with the pulse duration and the impact parameter, we consider the angle-resolved azimuthal contrast defined by

$$C_{tot,b_x}(\theta_f, \varphi_f) = \frac{\tilde{P}_{b_x}(\theta_f, \varphi_f)}{\int_0^{2\pi} d\varphi_f \, \tilde{P}_{b_x}(\theta_f, \varphi_f)/2\pi} - 1, \tag{9}$$

at a given impact parameter $b_x$. The contrast $C_{tot,b_x}(\theta_f, \varphi_f)$ vanishes when the signal at a given $\varphi_f$ equals that averaged over all $\varphi_f$. A positive (negative) value represents that the signal is larger (smaller) than the average. Figure 1(d) shows $C_{tot,b_x}(\theta_f, \varphi_f)$ for the two wave packet types with $\tau$ =1 fs and $b_x = 2$ Å. Due to the symmetry of the system, the contrast is totally symmetric with respect to the $x$-axis (i.e., vertical axis). For both wave-packet types, we observe strong azimuthal contrasts. For the $|k|$-Gauss wave packet (left panel), a strong up-down (i.e., one-fold) asymmetry appears at small $\theta_f$ (< 5 deg., inside the innermost dotted circle) angles. In addition, we observe a clear two-fold asymmetry at the relatively large ($\theta_f \geq 5$ deg.) angles. The two-fold asymmetric contrast here attains positive (red color) values at $\frac{\pi}{4} < \varphi_f < \frac{3\pi}{4}$ (right), and $\frac{5\pi}{4} < \varphi_f < \frac{7\pi}{4}$ (left) and negative (blue color) values at $\frac{7\pi}{4} < \varphi_f < 2\pi$, $0 < \varphi_f < \frac{\pi}{4}$ (up) and $\frac{3\pi}{4} < \varphi_f < \frac{5\pi}{4}$ (down). For the $k_{\parallel}$-Gauss wave packet (right panel), the two-fold asymmetry [i.e., blue lobes at $\varphi_f = \frac{\pi}{2}, \frac{3\pi}{2}$ (right, left) and red lobes at $\varphi_f = 0, \pi$ (up, down)] also appears but the sign is opposite compared to the $|k|$-Gauss. The asymmetry at small $\theta_f$ is much weaker than in the case of the $|k|$-Gauss. At large angles, there are also weak but non-negligible up-down asymmetries. Contrasts at lower halves tend to be positive (red) while those at upper halves being negative (blue).

To quantitatively discuss the azimuthal asymmetries and their dependence on $b_x$ and $\tau$, we introduce the following quantities [40],

$$A_1(b_x) = \int \sin\theta_f \, d\theta_f \frac{\int_0^{\pi/2} d\varphi_f \, \tilde{P}_{b_x} + \int_{3\pi/2}^{2\pi} d\varphi_f \, \tilde{P}_{b_x} - \int_{\pi/2}^{3\pi/2} d\varphi_f \, \tilde{P}_{b_x}}{\int_0^{2\pi} d\varphi_f \, \tilde{P}_{b_x}}, \tag{10}$$

$$A_2(b_x) = \int \sin\theta_f \, d\theta_f \frac{\int_0^{\pi/4} d\varphi_f \tilde{P}_{b_x} + \int_{3\pi/4}^{5\pi/4} d\varphi_f \tilde{P}_{b_x} + \int_{7\pi/4}^{2\pi} d\varphi_f \tilde{P}_{b_x} - \int_{\frac{\pi}{4}}^{\frac{3\pi}{4}} d\varphi_f \tilde{P}_{b_x} - \int_{5\pi/4}^{7\pi/4} d\varphi_f \tilde{P}_{b_x}}{\int_0^{2\pi} d\varphi_f \tilde{P}_{b_x}}, \tag{11}$$

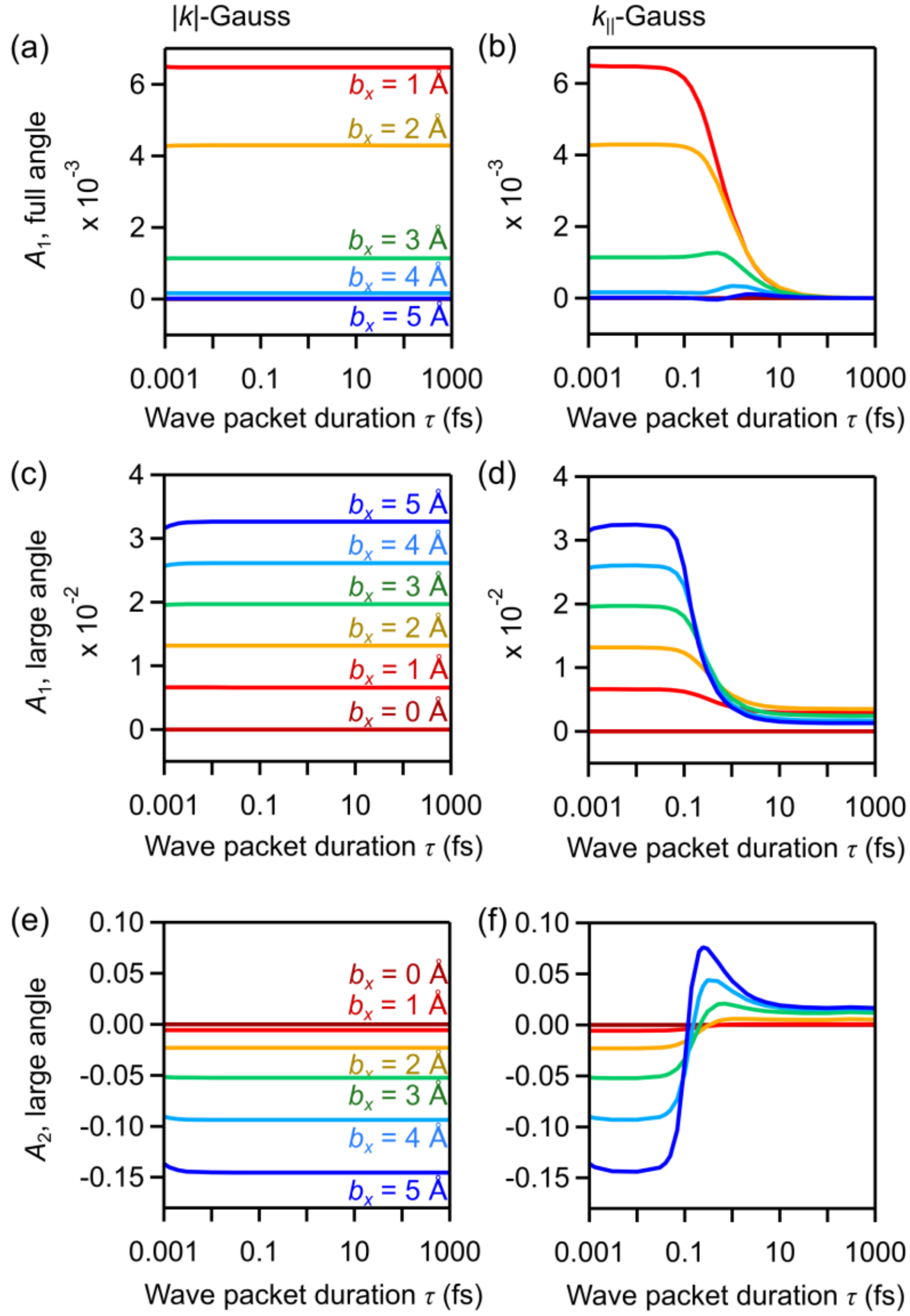

Fig 3. Azimuthal asymmetries for the $|k|$-Gauss wave packet (left panels) and those for the $k_\parallel$-Gauss wave packet (right panels), respectively. (a),(b) One-fold azimuthal asymmetries. See Eq. (10) for the definition. (c),(d) One-fold azimuthal asymmetries at the large ($\theta_f \geq 5^\circ$) final angles. (e),(f) Two-fold azimuthal asymmetries at the large final angles. See Eq. (11) for the definition. The two-fold azimuthal asymmetries for the entire final angle range were not plotted since their magnitudes are very small (<0.01%). When $b_x$ is negative, the sign of the one-fold azimuthal asymmetries is inverted, but the sign of the two-fold azimuthal asymmetries remains the same [40].

where the arguments of $\tilde{P}_{b_x}(\theta_f, \varphi_f)$ are omitted for notation convenience. These quantities attain single values at a given $b_x$ and $\tau$. A positive (negative) value of $A_1(b_x)$ means that signals are stronger (weaker) in the direction of the target shift. On the other hand, a positive (negative) value of $A_2(b_x)$ means that signals along *x* axis [vertical direction in Figs. 1(c), (d)] are stronger (weaker) than that along *y* axis (horizontal). As has been shown in [40] for the |*k*|-Gauss wave packet, the sign of $A_1(b_x)$ is reversed by the reversal of the sign of $b_x$, but the sign of $A_2(b_x)$ is not. Figures 3(a) and (b) show $A_1(b_x)$ for the two wave-packet types for various $b_x$ and $\tau$. Integrals were performed over the angular range of $0^\circ \leq \theta_f \leq 15^\circ$. As in the total scattering probabilities [Figs. 2(a)], $A_1(b_x)$ of the |*k*|-Gauss wave packet [Fig. 3(a)] is nearly independent on the duration $\tau$. However, $A_1(b_x)$ of the $k_{\|}$-Gauss wave packet [Fig. 3(b)] is strongly affected by $\tau$. At very short durations ($\tau$ < 0.1 fs), the value of $A_1(b_x)$ is constant and almost the same as those of the |*k*|-Gauss wave packet. However, $A_1(b_x)$ decays with $\tau$ at the range of 0.1 fs < $\tau$ < 10 fs and is almost zero at $\tau$ > 10 fs. The magnitude of $A_1(b_x)$ is maximal at $b_x = 1$ Å for both cases, which is consistent with the centre-of-mass analysis often used in the atomic-resolution STEM [12]. The dependence on $b_x$ represents the spatial resolution of the STEM, and the value of $b_x = 1$ Å is given by the spot size of the electron beam (1 Å rms in this work) [40]. We conducted the same analysis for $A_2(b_x)$, but the magnitudes are smaller than $10^{-4}$ at any $b_x$ and $\tau$ for both cases, and thus they are not plotted. The small magnitude is due to the strong signals at small $\theta_f$ angles [< 5 deg., inside the innermost dotted circle in Fig. 1(c)] where the two-fold azimuthal asymmetries are hardly seen [Fig. 1(d)].

In contrast to signals at small final $\theta_f$ angles, clearer two-fold azimuthal asymmetries are seen at large angles ($\theta_f \geq 5^\circ$) [Fig. 1(d)] where the scattered part $I_{sc,\boldsymbol{b}}(\boldsymbol{k}_f)$ [Eq. (5)] is dominant. We therefore evaluate and plot $A_1(b_x)$ in Figs. 3(c)-(d) and $A_2(b_x)$ in Figs. 3(e)-(f) calculated over the angular range of $5^\circ \leq \theta_f \leq 15^\circ$ in Eqs. (10) and (11). This situation corresponds to the use of an annular dark field detector. The magnitudes of both the one-fold and two-fold asymmetries increase with $b_x$, see also [40]. Magnitudes of $A_2(b_x)$ now exceeds 10% at $b_x = 5$ Å [Figs. 3(e)-(f)]. The asymmetries of the |*k*|-Gauss wave packet [Figs. 3(c) and (e)] are, again, nearly independent of the wave-packet duration $\tau$. On the other hand, the asymmetries of the $k_{\|}$-Gauss wave packet [Figs. 3(d) and (f)] show drastic changes at the range of 0.1 fs < $\tau$ < 10 fs. The magnitudes of $A_1(b_x)$ [Fig. 3(d)] decrease but do not converge to zero at large durations. The values of $A_2(b_x)$ [Fig. 3(f)] are negative at very short durations and the sign changes to positive at $\tau \sim 0.1$ fs. They become maximal at $\tau \sim 0.2$ fs and decay to non-zero positive values at large durations.

The above results show that the total scattering probabilities as well as the angular asymmetries can be controlled by the type of wave packets, and the strong sensitivity to the duration can be induced with a proper choice of the wave-packet type (here $k_{\|}$-Gauss wave packet). In the next section, we will discuss the origin of the above findings.

## 4. Discussion

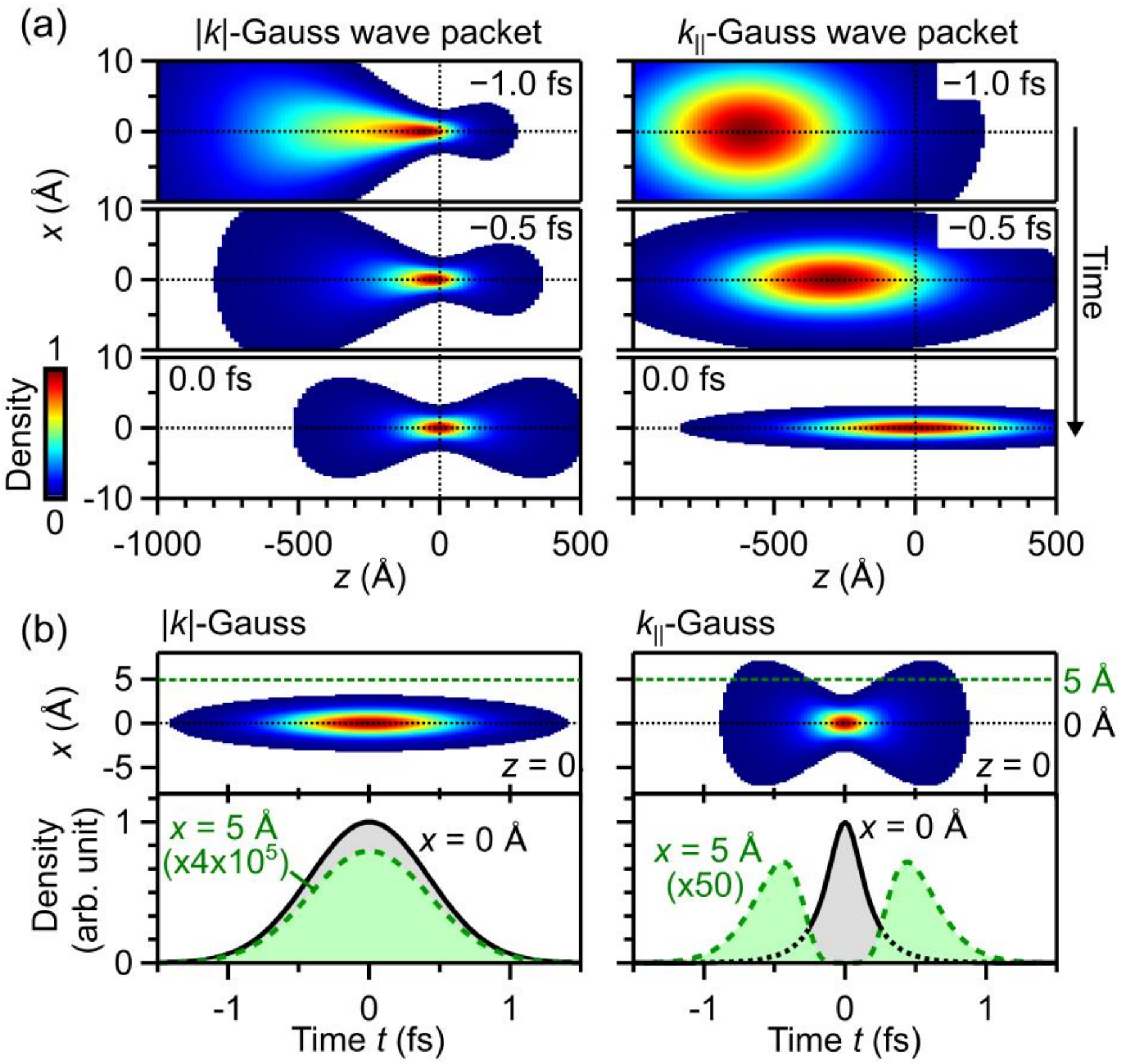

Fig 4. Sub-fs real-space dynamics of the electron wave packets with $\tau = 1$ fs. (a) Snapshots of the wave packets on the $xz$ plane. At time zero (lowest panels), the wave packets are transversally focused. The result of each panel is normalized independently. (b) Upper panels: temporal evolution of the probability densities at $z = 0$. Lower panels: Slices at $x = 0$ and 5 Å.

To gain a physical understanding of these findings, we first consider the spatial distributions of the wave packets and relate them to the total elastic scattering probabilities (Fig. 2). Figure 4(a) shows snapshots of the probability densities $|\psi_e(x, y = 0, z, t)|^2$ in the $xz$ plane for a duration of $\tau = 1$ fs, evolving on sub-fs time scales. At $t = 0$ fs (lowest panels), the wave packets are spatially focused. The upper panels show the real-space densities at negative times, i.e., before the focus. The overall density distribution of the $|k|$-Gauss wave packet (left panels) follows the shape of a focused Gaussian beam known from optics [45] regardless of the wave-packet duration $\tau$, suggesting the independence of the scattering probabilities on $\tau$ (left panels of Fig. 2). Figure 4(b) shows the temporal evolution of the wave-packet densities at $z = 0$ [vertical slices of Fig. 4(a)], where the target atom is placed. The probability densities are peaked at around $x = 0$. This leads to the observed significant decay of the scattering probability with the impact parameter.

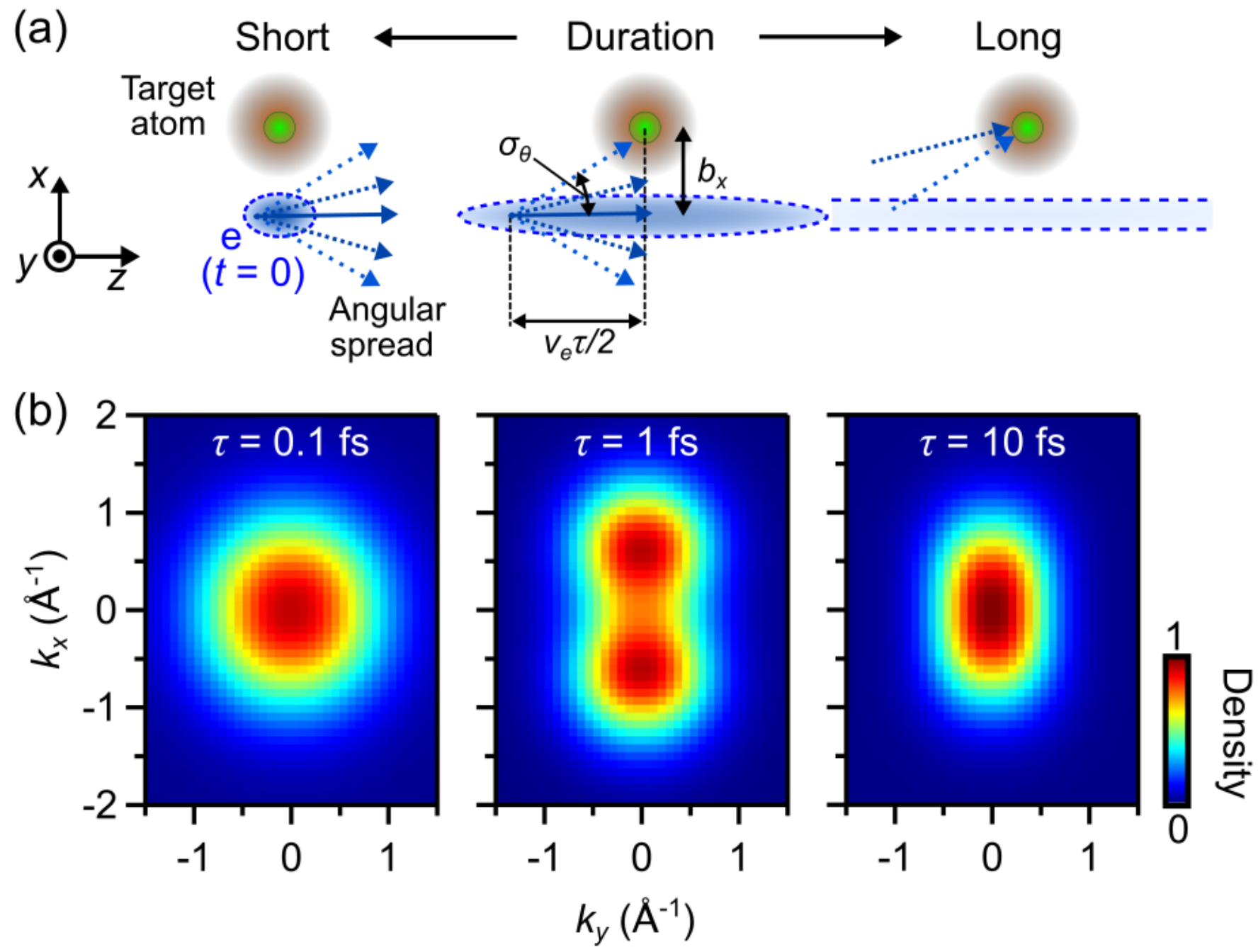

Fig 5. Spatial and momentum overlaps of the $k_{||}$-Gauss wave packet with a target. (a) Illustration of the wave-packet duration dependence. The FWHM wave-packet longitudinal size is $v_e\tau$. (b) Momentum distribution of the wave packet at the position of the target, given by the Gabor transform [Eqs. (13)-(14)].

On the other hand, the real-space densities of the $k_{||}$-Gauss wave packet [right panels of Fig. 4(a)] have a two-dimensional Gaussian shape at any time. Interestingly, we find non-negligible densities at large $x$ and $z = 0$ (vertical lines), at −1.0 fs and −0.5 fs. The right panel of Fig. 4(b) shows the temporal evolution of the densities at $z = 0$. Noteworthy, at a large $x$, there are two density peaks in time, separated by 0.9 fs at $x = 5$ Å (green curve in lower panel), whose potential applications will be briefly discussed in the concluding remarks. Because the longitudinal size is determined by the wave-packet duration $\tau$, the overlap with the target at larger $x$ is strongly affected by $\tau$. The illustrations in Fig. 5(a) qualitatively explain the dependence on $\tau$. The blue elliptical circles show the spatial distribution of the wave packets at $t = 0$. From each point of the wave packet, the electron spreads with a given divergence angle (arrows). At a very short duration (left), even a large spreading angle component does not cross the target atom. At an intermediate duration (middle), a large-angle component starting from the edge of the distribution traverses the target. At a very long duration (right), components from small to large angles can cross the target but the contribution of each component is small because the wave packet is extended along $z$. Accordingly, the scattering probability decreases with ~$1/\tau$ [right panel of Fig. 2(a)]. The wave-packet duration corresponding to the intermediate regime for $b_x = 5$ Å can be estimated as $b_x/(\sigma_\theta v_e/2)$ ~ 0.8 fs, see the middle panel of Fig. 5(a).

To quantitively confirm the above discussion, we simulated time-integrated probability densities at $y = z = 0$, that is $\int_{-\infty}^{+\infty} |\psi_e(x, y = 0, z = 0, t)|^2\, dt$, for the two types of wave packets. The scattering probability is approximately given by

$$P_{model}(x) = \frac{\sigma_{elastic} \int_{-\infty}^{+\infty} |\psi_e(x, y = 0, z = 0, t)|^2\, dt}{2\pi \int_0^{+\infty} x dx \int_{-\infty}^{+\infty} |\psi_e(x, y = 0, z = 0, t)|^2\, dt}, \tag{12}$$

where $\sigma_{elastic}$ is the total elastic cross section and is 0.00288 Å$^2$ for a hydrogen atom at 10-keV energy [42]. The denominator is the total flux passing through the $z = 0$ plane. The numerator approximately gives the flux passing through a circle at $x = b_x$ whose area is equal to the cross section $\sigma_{elastic}$. Circles in Fig. 2(a) show the calculated $P_{model}(x = b_x)$. We obtain good matches between the probability densities (circles) and the total scattering probabilities (curves) for both wave-packet types. Even the curves with complicated shapes for the $k_{||}$-Gauss wave packet (right panel) are reproduced well. The good agreement validates the physical picture of Figs. 4-5 and confirms the connection between the local probability density and the scattering probability. At points of very small ($<10^{-5}$) scattering probabilities, the local densities underestimate the former. This discrepancy could originate from the long-range nature of the Coulomb potential of the target atom.

The physical picture in Fig. 5(a) also explains the positive two-fold azimuthal asymmetries $A_2(b_x) > 0$ at $\tau$>10 fs for the $k_{||}$-Gauss wave packet and $|b_x| > 0$, seen in Fig. 3(f). Intuitively, components with transversal momenta along $x$ can likely reach the target compared to those along $y$ (perpendicular to the plane). This can produce an anisotropy in the ($k_x$,$k_y$) distribution of electrons coming to the target. Because the scattering probability is higher with smaller momentum transfer $\left|\hbar(\boldsymbol{k}_f - \boldsymbol{k}_i)\right|$, i.e., the scattering probability is the strongest at forward direction, this anisotropy is translated into the angular distribution of scattered electrons, leading to the azimuthal asymmetry. We quantify the momentum anisotropy by using the Gabor transform as

$$Q(x, y, z, k_x, k_y) = \iint dt dk_z |G(x, y, z, k_x, k_y, k_z, t)|^2\,, \tag{13}$$

with

$$G(x, y, z, k_x, k_y, k_z, t) = \frac{1}{(2\pi)^{\frac{3}{2}}} \iiint \psi_e(x', y', z', t) \exp\left(-\frac{(x'-x)^2}{2\delta_x^2}\right) \exp\left(-\frac{(y'-y)^2}{2\delta_y^2}\right) \exp\left(-\frac{(z'-z)^2}{2\delta_z^2}\right)$$
$$\times \exp(-\mathrm{i}k_x x') \exp(-\mathrm{i}k_y y') \exp(-\mathrm{i}k_z z')\, dx' dy' dz', \tag{14}$$

where $\delta_x$, $\delta_y$, $\delta_z$ are the rms spatial width of interests. Equations (13) and (14) show that the Gabor transform extracts information on the momentum density distributions $(k_x, k_y, k_z)$ of the wave packet at the position of the target $(x, y, z)$. Figure 5(b) shows the two-dimensional momentum distribution $Q(x, y, z, k_x, k_y)$ of the $k_{||}$-Gauss wave packet at $x$ = 5 Å, $y$ = $z$ = 0 Å for packet durations $\tau$ = 0.1 fs (left), 1 fs (middle) and 10 fs (right). We chose $\delta_x = \delta_y = \delta_z = 2$ Å to see their differences clearly. At the short duration of 0.1 fs, the momentum distribution is isotropic. On the other hand, at $\tau$ =1 fs, the distribution is significantly anisotropic and there are two peaks at $k_x \sim \pm 0.7$ Å$^{-1}$. At $\tau$ =10 fs, the $k_x$-$k_y$ anisotropy is reduced but not lost. This trend matches with the physical picture of Fig. 5(a) and explains the results in Fig. 3(f), where $A_2(b_x)$ peaks at hundreds of attoseconds and decreases at longer durations. Based on these results, we conclude that the positive $A_2(b_x)$, i.e., stronger scattering probabilities towards/against the target direction ($\varphi_f = 0, \pi$), for the $k_{||}$-Gauss wave packet is attributed to the anisotropic momentum distribution given by the real-space wave-packet dynamics on few-fs time scales.

Next, we explain the origin of the negative asymmetry $A_2(b_x) < 0$ appearing in the large-angle (i.e., dark field) elastic scattering of the $|k|$-Gauss wave packet at any $\tau$ and the short-duration $k_{||}$-Gauss wave packet [see Figs. 3(e)-(f)]. We return to Eq. (5) and focus on the term $e^{\mathrm{i}k_f(\hat{\boldsymbol{k}}_i - \hat{\boldsymbol{k}}_i')\cdot\boldsymbol{b}}$ that follows from evaluating the norm squared, see Eq. (15) below. The physical origin of this phase term is illustrated in Fig. 1(b). For an electron with an incident angle $\theta_i$ on the $xz$ plane, one can find the difference in the geometrical path length as compared to the case of $b_x = 0$, which is given by $\hat{\boldsymbol{k}}_i \cdot \boldsymbol{b} = b_x \sin\theta_i$, shown by the red arrow. Note that a similar discussion is applied to electron diffraction by molecules or crystals. The scattering probability in Eq. (5) contains the relative phase for $\boldsymbol{k}_i$ and $\boldsymbol{k}_i'$, that is $e^{\mathrm{i}k_f(\hat{\boldsymbol{k}}_i - \hat{\boldsymbol{k}}_i')\cdot\boldsymbol{b}}$. Because smaller momentum transfer $|\hbar k_i(\hat{\boldsymbol{k}}_i - \hat{\boldsymbol{k}}_f)|$ leads to higher scattering probabilities, the scattering to $\varphi_f$ = 0 and π can be represented by that occurring on the $xz$ plane, in which the phase term is given by $e^{\mathrm{i}k_f(\sin\theta_i - \sin\theta_i')\cdot b_x}$, while the scattering to $\varphi_f$= π/2 and 3π/2 can be represented by an event on the $yz$ plane, in which the phase term is $e^0$ = 1 because $\hat{\boldsymbol{k}}_i$ and $\hat{\boldsymbol{k}}_i'$ are perpendicular to $\boldsymbol{b}$. Therefore, the term inside the integral of Eq. (5) is added in phase for the scattering on the $yz$ plane, leading to the higher scattering probabilities at $\varphi_f$ = π/2, 3π/2 than at $\varphi_f$ = 0, π. The $\boldsymbol{b}$-dependent phases discussed here are common for any target species, and therefore the two-fold asymmetry does not strongly depend on the specific type of target atom, see also the comparison performed for three different target atoms in reference [40].

Finally, we discuss the origins of the $\tau$-dependences of the one-fold azimuthal asymmetries $A_1(b_x)$ in Figs. 3(a)-(d). Because the un-scattered part $I_{unsc}(\boldsymbol{k}_f)$ [Eq. (3)] is symmetric, the azimuthal asymmetry at the entire angular range is dominated by the interference part $I_{int,\boldsymbol{b}}(\boldsymbol{k}_f)$ [Eq. (4)], while that at large angles is given by the scattered part $I_{sc,\boldsymbol{b}}(\boldsymbol{k}_f)$ [Eq. (5)]. As in Eq. (4), $I_{int,\boldsymbol{b}}(\boldsymbol{k}_f)$ is caused by the interference between $a_e(\boldsymbol{k}_f)$ and $a_e(k_f\widehat{\boldsymbol{k}}_i)e^{-\mathrm{i}(\boldsymbol{k}_f-k_f\widehat{\boldsymbol{k}}_i)\cdot\boldsymbol{b}}f_{elas}(k_f\widehat{\boldsymbol{k}}_i,\boldsymbol{k}_f)$. The information on the target atom is encoded into the magnitude of the asymmetry via the elastic scattering amplitude $f_{elas}(k_f\widehat{\boldsymbol{k}}_i,\boldsymbol{k}_f)$. Because the kinetic energy is conserved by elastic scattering, the transition of $\boldsymbol{k}_i\to\boldsymbol{k}_f$ corresponds to the vertical shifts in the energy-angle picture in Fig. 1(e). In the case of the $|k|$-Gauss wave packet whose momentum-space density is a two-dimensional Gaussian [left panel of Fig. 1(e)], the increase of $\tau$ solely reduces the horizontal (i.e., energy) width. Because the angular ($\theta_i$) distribution is independent on $\tau$, the asymmetry does not change with $\tau$ [Fig. 3(a)]. In contrast, the density of the $k_{||}$-Gauss wave packet having an energy-angle entanglement shows an arc shape [right panel of Fig. 1(e)]. With the increase of $\tau$, the arc becomes sharper and the overlap between $a_e(\boldsymbol{k}_f)$ and $a_e(k_f\widehat{\boldsymbol{k}}_i)e^{-\mathrm{i}(\boldsymbol{k}_f-k_f\widehat{\boldsymbol{k}}_i)\cdot\boldsymbol{b}}f_{elas}(k_f\widehat{\boldsymbol{k}}_i,\boldsymbol{k}_f)$ becomes smaller. Consequently, the magnitude of the $A_1(b_x)$ asymmetry decreases with $\tau$ [Fig. 3(b)].

A similar argument holds for $A_1(b_x)$ at large angles [Fig. 3(c)-(d)] where the scattered part [Eq. (5)] is dominant. Equation (5) can be re-written as

$$I_{sc,\boldsymbol{b}}(\boldsymbol{k}_f)=\frac{k_f^2}{4\pi^2}\int d\widehat{\boldsymbol{k}}_i'\int d\widehat{\boldsymbol{k}}_i\, a_e^*(k_f\widehat{\boldsymbol{k}}_i')a_e(k_f\widehat{\boldsymbol{k}}_i)e^{\mathrm{i}k_f\boldsymbol{b}(\widehat{\boldsymbol{k}}_i-\widehat{\boldsymbol{k}}_i')}f_{elas}^*(k_f\widehat{\boldsymbol{k}}_i',\boldsymbol{k}_f)f_{elas}(k_f\widehat{\boldsymbol{k}}_i,\boldsymbol{k}_f). \qquad (15)$$

This equation shows that the scattered part is given by the coherent contribution of the two paths that are $\boldsymbol{k}_i'\to\boldsymbol{k}_f$ and $\boldsymbol{k}_i\to\boldsymbol{k}_f$ where $|\boldsymbol{k}_i'|=|\boldsymbol{k}_i|$. The magnitude of $A_1(b_x)$ at large angles therefore depends on the difference in the phases of $f_{elas}^*(k_f\widehat{\boldsymbol{k}}_i',\boldsymbol{k}_f)$ and $f_{elas}(k_f\widehat{\boldsymbol{k}}_i,\boldsymbol{k}_f)$. Accordingly, the one-fold asymmetry at large angles is sensitive to the phase shift due to the potential of the target atom [40]. For the case of the $k_{||}$-Gauss wave packet [Fig. 3(d)], the increase of $\tau$ reduces the range of $\theta_i$ at an energy, and thus the contribution starting from the different initial angles ($\widehat{\boldsymbol{k}}_i'$ and $\widehat{\boldsymbol{k}}_i$) and reaching the same final angle $\widehat{\boldsymbol{k}}_f$ becomes weak. However, even at the limit of $\tau\to+\infty$, there are still couplings between different initial azimuthal angles ($\varphi_i$). Indeed, in our previous study [40], we have shown that the azimuthal asymmetries appear even with a fixed $\theta_i$. Accordingly, the magnitude of the azimuthal asymmetries for the $k_{||}$-Gauss wave packet decreases with $\tau$ but does not converge to zero at the limit of $\tau\to+\infty$, except for $b_x=0$. As above, the large-angle $A_1(b_x)$ for the $|k|$-Gauss wave packet [Fig. 3(c)] is nearly independent on $\tau$ because of the initial angular distribution independent on $\tau$.

Although we have limited our discussion to elastic scattering in this work, similar considerations can be made for inelastic processes. Equation (5) can readily be applied to inelastic scatterings by replacing the elastic scattering amplitude $f_{elas}(k_f\widehat{\boldsymbol{k}}_i,\boldsymbol{k}_f)$ with that for the inelastic process, and a discussion of the large-angle angular asymmetries following the one presented here could be made. In contrast, for the interference term [Eq. (4)], which dominates the angular asymmetries at small angles, to be non-vanishing, the un-scattered and scattered paths need to reach the same final state. This is difficult to achieve with a monochromatic electron beam and a steady-state target. However, it is possible with a target in a superposition state and with an ultrashort electron wave-packet having a broad energy spectrum, and hence, interference between inelastic and elastic scatterings can occur under such conditions. Also, applications to a molecular or a solid-state target are possible by replacing the atomic scattering amplitude $f_{elas}\,(k_f\hat{\boldsymbol{k}}_i, \boldsymbol{k}_f)$ in Eqs. (4)-(5) by that for a more complex target.

## 5. Experimental feasibility

Before concluding, we discuss possible experimental schemes based on existing techniques for obtaining the considered wave packets. A sub-picosecond electron pulse can be obtained with a photocathode gun [46]. To achieve packet durations $\tau$ below 100 fs, which cannot be directly obtained from ultrashort electron guns [46], a buncher has to be employed, in which a light-wave-driven [green curve in Fig. 6(a)] energy modulation is applied to the incident electron pulse. Examples of the buncher include a laser-excited thin membrane [17,18,25,47–51], a dielectric laser accelerator [22–24] or a two-color ponderomotive potential [21]. A microwave-driven buncher is also applicable with careful management of the temporal jitter [52]. The buncher imposes spatially homogeneous but temporally varying acceleration and deceleration to electrons following the oscillation of the light wave. The energy modulation at the buncher evolves into the temporal density modulation (i.e., temporal compression) of the electron pulse during the subsequent free-space propagation [26] as the accelerated component catches up with the decelerated component in front. The buncher-target distance or the amplitude of the light wave is chosen such that the wave packet duration is the shortest at the position of the target. When the incident electron wave packet has a Gaussian temporal profile and is bunched using a linear part (i.e., near the zero-crossing) of the light-wave or the microwave, the energy spectrum of the modulated wave packet is also Gaussian. The $|k|$-Gauss wave packet has the same energy spectrum over the incident angle $\theta_i$, and therefore, can be produced by focusing using a magnetic lens (Fig. 6) with a proper choice of the electron source and focusing geometry [53]. On the other hand, the $k_{||}$-Gauss wave packet has a constant longitudinal momentum ($k_{\parallel} = k_i \cos\theta_i$) distribution over $\theta_i$, which corresponds to an energy spectrum whose mean energy increases with the angle $\theta_i$. To generate such a wave packet, we propose to employ another energy modulator of a thin membrane excited by long-wavelength light such as a THz wave (red curve in Fig. 6) [15,19] whose cycle is longer than the wave packet duration such that a temporally uniform energy shift (deceleration, here) can be achieved. A smaller incident angle with respect to the electron beam is preferred to induce an axially symmetric modulation (i.e., circular field distribution) with minor astigmatism control. By setting the electron beam diameter slightly larger than the focused light, electrons experience the strongest deceleration at the beam center and reduced deceleration with increased distance from the center. Subsequent spatial focusing by a lens produces the $k_{||}$-Gauss-like wave packet. Alternatively, the space-energy coupled wave packet could also be obtained with a dielectric laser accelerator which relies on the spatially-decaying optical near fields and intrinsically provides position-dependent energy modulation amplitude [22–24]. Verification of these schemes with a full numerical simulation deserves an entire study on its own. We note that the capability of space-time energy control with light waves can produce electron wave packets with unique properties that are markedly different than what can be obtained by manipulating laser pulses (i.e., light wave packets).

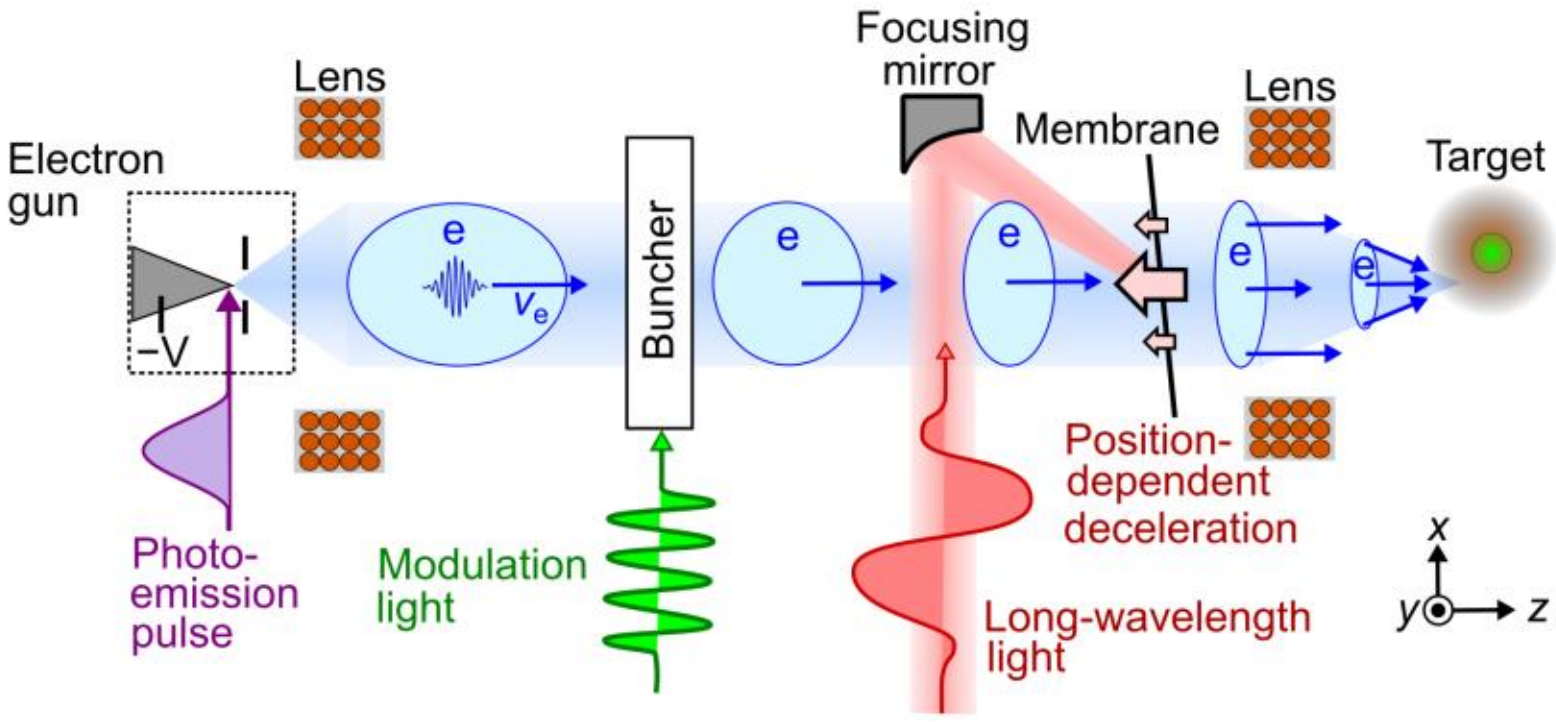

Fig. 6. Proposed scheme for producing $|k|$-Gauss and $k_{||}$-Gauss wave packets. See text for details.

In addition, we briefly estimate the orders of magnitudes of the electron-beam flux required for observing the predicted azimuthal asymmetries. By considering the shot-noise of $\sqrt{N}$ with the total number of electrons $N$ detected, the $A_1(b_x)$ asymmetry [Figs. 3(a),(b)] of the orders of $10^{-3}$ can be detected with $N = 10^6$. The $A_2(b_x)$ asymmetry at large angles [Figs. 3(e),(f)], for example, 5% at $b_x = 2$ Å, can be detected with $N = 10^8$-$10^9$ when the total scattering probabilities in Fig. 2 are considered. These amounts of flux are feasible with ultrafast electron microscopes. For example, 100-sec integration is enough with a single-electron pulse at 1 MHz repetition rate. Since these azimuthal asymmetries are universal to any atomic elements as shown in [40], the estimated integration time will be much shorter with a heavy-element target.

## 6. Conclusion

In summary, motivated by the rapid advances on light-driven control of electron beams, we investigated the fundamental aspects of elastic scattering of electron wave packets shaped in 3D space. The dependence of the scattering probabilities on the wave-packet type, duration and impact parameters were elucidated by the spatial-momentum overlap between the incident beam with the target atom occurring on the few-fs time scale. The two-fold azimuthal asymmetry induced by the displacement of the target from the electron beam center is attributed to two competing effects, the anisotropy of momentum distribution, which was revealed by the Gabor transform, and the geometrical phase determined by the impact parameter. Both the scattering probability and asymmetry related to the use of the wave packet with a Gaussian distribution in the longitudinal momentum strongly depend on its duration in the range of 0.01-100 fs, suggesting a novel *in-situ* approach to characterize the ultrashort wave packet in an extreme temporal parameter range purely from time-unresolved scattering data. Moreover, for certain shaped electron pulses, the target displaced from the beam center experiences two electron bursts within a single wave packet [right panel of Fig. 4(b)] whose separation is tunable with the displacement distance on the few-fs time scale level (0.5 fs at $b_x$ = 3 Å and 0.9 fs at $b_x$ = 5 Å), which could be useful for coherently exciting the target through the free-electron—bound electron transition scheme [27,28], probing the electronic coherence of the atom [30,54], or for an electron-pump—electron-probe experiment of sub-fs carrier dynamics. These aspects can be explored by extending our framework to inelastic processes as well as scattering by a target in a coherent superposition of states.

**ACKNOWLEDGEMENTS**

This work is supported by the Japan Science and Technology Agency (JPMJAX21AO and JPMJFR2228), the MEXT/JSPS KAKENHI JP21K21344, the Research Foundation for Opto-Science and Technology, JSPS Invitational Fellowship for Research in Japan, the Gordon and Betty Moore Foundation (GBMF) through Grant No. GBMF4744 and GMBF11473, ERC Advanced Grant No. 884217 “AccelOnChip” and the Independent Research Fund Denmark (Grant ID: 10.46540/4283-00004B) .